\documentclass[onecollarge,natbib]{svjour2}
\bibpunct{[}{]}{;}{n}{}{,} % to get "[numbered]" references from natbib
\smartqed  % flush right qed marks, e.g. at end of proof
\usepackage{graphicx}
\usepackage{bm}
\journalname{Few-Body Systems}
\begin{document}

\title{Gauge invariance, Lorentz covariance and canonical quantization in nucleon structure studies}


\author{Fan Wang, X.S. Chen, W.M. Sun, P.M. Zhang and C.W. Wong }


\institute{Fan Wang \at Department of Physics, Nanjing University, Nanjing, 210093, China \\
              Tel.: 86-25-83592325, Fax:  86-25-83326028 \\
              \email{fgwang@chenwang.nju.edu.cn}}
           %\and
           %X.S. Chen \at School of Physics, Huazhong university of Science and technology, Wuhan, 430074, China
           %\\ \email{cxs@hust.edu.cn}
           %\and
           %W.M. Sun \at Department of Physics, Nanjing University, Nanjing, 210093, China \\
           %\email{sunwm@chenwang.nju.edu.cn}
           %\and
           %P.M. Zhang \at Institute of Modern Physics, Chinese Academy of Science, Lanzhou, 730000, China \\
           %\email{zhpm@impcas.ac.cn}
           %\and
           %C.W. Wong \at Department of Physics, UCLA, Los Angeles, CA90024, USA
           %\email{cwong@physics.ucla.edu}

\date{Received: date / Accepted: date}

\maketitle

\begin{abstract}
There are different operators of quark and gluon momenta, orbital angular momenta, and gluon spin
in the nucleon structure study. The precise meaning of these operators are studied
based on gauge invariance, Lorentz covariance and canonical quantization rule. The advantage
and disadvantage of different definitions are analyzed. A gauge invariant canonical
decomposition of the total momentum and angular momentum into quark and gluon parts is
suggested based on the decomposition of the gauge potential into gauge invariant (covariant)
physical part and gauge dependent pure gauge part. Challenges to this proposal are answered.

\keywords{Physical and pure gauge potentials; Gauge invariant canonical quark and gluon momenta,
orbital angular momenta and spins; Homogeneous and non-homogeneous Lorentz transformations; Gauge invariant
decomposition and gauge invariant extension; Classical and quantum measurements.}
\end{abstract}

\section{Introduction}
\label{intro}

After more than a quarter century of measurements by different groups, the quark spin contribution to the
nucleon spin has been found to be only about one third of its known value, while the contribution from
polarized gluons seems to be quite small. Attention therefore turns to the contributions from the orbital
motion of quarks, and the spin and orbital motion of antiquarks and gluons. It is then relevant to ask if
the spin and orbital angular momentum (OAM) of gluons inside nucleons are separately measurable.

The first issue is gauge invariance. This means in QED that calculations using different vector
potentials that generate the same electromagnetic (em) fields give the same physical results. Many
formulas describing the motion of electrons in em fields appear at first sight to be gauge-dependent,
with non-unique possible values for the vector potential used, and hence of questionable
measurability~\cite{ref1}. Yet the momentum and OAM of electrons moving in em fields, and especially
the energies of atomic electrons have all been successfully measured experimentally at a time when
the corresponding theoretical descriptions were not explicitly gauge invariant. These measurements
owe their success to a hidden or implicit gauge invariance in the formulas used.
We shall describe, in Section 2, how hidden gauge invariance arises not just in free space and in QED
in the Coulomb gauge~\cite{ref2}, but also in other members of a certain family of gauges including the
Lorenz gauge.

The confusion concerning gauge invariance and measurability would not have appeared if all formulas
are first made {\it explicitly} gauge-independent. Explicitly gauge invariant expressions for the
momentum and for the decomposition of the total AM of a gauge boson into its spin and OAM parts seem to
be unknown~\cite{ref3} before our proposals for them. We shall review our proposals, and a few other
related proposals in Section 3.

A number of objections have been raised against our operators by X. Ji~\cite{ref4}. We list some of them
below, together with our response to each in the nutshell drawn as parentheses:
\begin{enumerate}
\item[(1)]
Our momentum operator does not contain the gauge interaction. (It should not.)
\item[(2)]
Our decomposition is not Lorentz covariant. (It is Lorentz covariant.)
\item[(3)]
Our operators are nonlocal, and therefore unacceptable. (They are nonlocal but measurable.)
\end{enumerate}
In the final Section 4, we shall explain our responses in more detail and discuss a number of related issues.

This short review gives only a schematic analysis of these issues. More details will be given elsewhere.

\section{Hidden or implicit gauge invariance}
\label{sec:2}

In the standard Yang-Mills gauge theory, a physical gauge invariant interaction can be added to a free
Dirac equation by introducing a physical gauge field. We are here also interested in constructing a
gauge invariant theory that contains no physical interaction at all. The free Dirac equation for an
electron of charge $e$ then takes the gauge invariant form (in natural units $\hbar = c = 1$)
\begin{equation}
\left[i\gamma^{\mu}(\partial_{\mu} + ieA'_{\mu, {\rm pure}}) - m \right]\psi' = 0,
\end{equation}
where the ``pure gauge'' potential $A'_{\mu,pure}$ is introduced solely to cancel an arbitrary
local gauge transformation $U(x)$ added to a Dirac wave function $\psi(x)$
\begin{eqnarray}
\psi'(x) = U(x)\psi(x), \quad U(x) &=& e^{-ie\omega'(x)}: \nonumber \\
U^{-1}(x)\left[\partial_{\mu} + ieA'_{\mu, {\rm pure}}(x)\right]U(x) &=& \partial_{\mu}, \quad
{\rm if} \quad A'_{\mu, {\rm pure}}(x) = \partial_\mu \omega'(x).
\label{gaugeX}
\end{eqnarray}
Then $\psi(x)$ satisfies an unprimed version of Eq. (1) with $A_{\mu,{\rm pure}} = 0$, and is free of any
arbitrary local gauge function (i.e.,  $\omega(x) = 0$). A vector potential $A_{\mu,{\rm  pure}}$ if present is called pure gauge or nondynamical because 
it does not give rise to a physical gauge interaction:
\begin{eqnarray}
F^{\mu\nu}_{{\rm pure}}=\partial^{\mu}A^{\nu}_{{\rm pure}}-\partial^{\nu}A^{\mu}_{{\rm pure}} = 0.
\end{eqnarray}
The unprimed Dirac equation for $\psi(x)$, with the immaterial $A_{\mu,{\rm pure}} = 0$ left out, appears
at first sight to be gauge dependent, but it is not. Its gauge invariance is only hidden or implicit.

Suppose next that an em interaction is also present. Then the associated gauge potential will contain a 
physical part as well as a nonphysical pure gauge part:
$A_{\mu} = A_{\mu,{\rm pure}} + A_{\mu, {\rm phys}}$, where
\begin{eqnarray}
\vec{\partial}^2 A_{\mu, {\rm phys}}- \partial_\mu \partial_i A_{i, {\rm phys}} = \partial_i F_{i\mu}(x).
\label{Feq}
\end{eqnarray}
We are interested in the solution for $A_{\mu, {\rm phys}}$ satisfying the physical transverse wave condition
\begin{equation}
\partial_i A_{i,{\rm phys}}=0.
\label{transversality}
\end{equation}
This condition also defines the traditional Coulomb gauge, one of many choices of gauge that allows 
$A_{\mu}$ to be calculated uniquely from a given em field $F$. In the Coulomb gauge, the physical em 
field $F$ of a photon in free space reside in the 2D transverse ``physical subspace'' perpendicular to 
the photon momentum.

Gauge transformations cannot change the em fields in the photon's 2D physical space, by definition.
(Technically, this happens because $A_{{\rm phys}}$ commutes with $U(x)$. Such a condition is easily
satisfied in QED, but is harder to realize in non-Abelian gauge theories.)
They can only change $A_{\mu, {\rm pure}}$. That is,  $A_{\mu, {\rm pure}}$ carries the whole gauge
degree of freedom, while $A_{\mu, {\rm phys}}$ is gauge invariant:
\begin{equation}
A'_{\mu, {\rm phys}} = A_{\mu, {\rm phys}}, \quad
A'_{\mu,{\rm pure}} = A_{\mu, {\rm pure}} + \,\partial_{\mu}\omega'(x).
\end{equation}
Thus through a gauge transformation, one can eliminate the pure gauge part $A_{\mu, {\rm pure}}$
completely, while leaving the gauge invariant part $A_{\mu, {\rm phys}}$ intact to give Dirac's gauge
invariant result~\cite{ref5}. Formulas that show $A$ alone might still be gauge invariant, though
only implicitly.

Physically the simplest and most interesting $A_{\mu, {\rm pure}}$s for photons are those that do not
intrude into the photon physical subspace: For them $A_{\mu,{\rm pure}}$s have only $\parallel$ and 
timelike components in the 2D subspace orthogonal to the 
physical subspace. For all these gauges, $A_{{\rm phys}} = {\bf A}_\perp$. These gauge transformations 
include many gauge choices of physical interest, including the Lorenz gauge.

If one's emphasis is not on photons, other physically motivated subsidiary conditions different from 
Eq. (\ref{transversality}) might be more useful. The resulting $A_{\mu, {\rm phys}}$s are in general 
different from each other, and different from the photon quantity defined here. To each choice of 
physically motivated subsidiary conditions there is a family of pure gauge transformations that are 
nondynamical, with $F^{\mu\nu}_{{\rm pure}} = 0.$

\section{Gauge invariant decomposition of momentum and AM of a gauge system}
\label{sec:3}

To study the spin structure of nucleon, an $SU(3)$ color gauge field system, one wants to separate quark and
gluon spin and OAM contributions. Jaffe and Manohar (JM) first obtained such a decomposition~\cite{ref7},
\begin{equation}
\bf{J} = \int d^3 x \,\psi ^\dagger \frac 12 \bf{\Sigma} \psi + \int
d^3x \,\psi ^\dagger \bf{x} \times\frac 1i \bf{\nabla} \psi + \int d^3x \,\bf{E}\times \bf{A}
+\int d^3x \,E^i\bf{x}\times \bf{\nabla} A^i.
\label{JM}
\end{equation}
The advantage of this decomposition is that the individual terms all satisfy the $SU(2)$ AM algebra;
they are proper quark and gluon spins and OAMs. However, the terms not involving the quark spin
$\bf{\Sigma}$ are not gauge invariant.

 The gauge non-invariance originates from the quantization of the momentum and OAM of a
charged particle moving in an em field~\cite{ref1}. To obtain the Lorentz equation of a charged
particle moving in an em field, one employs the Lagrangian ${\cal L} = (1/2)mv^2 - e(A^0-\bf{v}\cdot\bf{A}).$
Following the standard canonical mechanics, one obtains the canonical momentum and OAM,
\begin{equation}
{\bf p} = m{\bf v} + e{\bf A},~~~~~\bf{L} = \bf{x}\times\bf{p}.
\end{equation}
Classically, they are gauge dependent and so are not measurable. Canonical quantization quantizes them as
canonical momentum and OAM no matter which gauge is used. Feynman had explained why we
quantize the canonical momentum {\bf p} rather than the mechanical momentum $m\bf{v}$~\cite{ref8}.
Canonical quantization appears to be gauge invariant. However, since the wave function of a charged
particle is still gauge dependent, the MEs of both canonical momentum and OAM remain gauge dependent.
That is, they are still not measurable.

 To remedy the gauge non-invariance of JM's decomposition, both our group and Ji obtained a gauge
invariant decomposition in 1997~\cite{ref9},
\begin{equation}
\bf{J} = \int d^3 \,x \psi ^\dagger \frac 12 \bf{\Sigma} \psi + \int
d^3x \,\psi ^\dagger \bf{x} \times\frac 1i \bf{D}\psi +\int
d^3x \,\bf{x}\times \left(\bf{E}\times\bf{B}\right),
\label{gid}
\end{equation}
where $\bf{D} = \bf{\nabla}-ie\bf{A}$. The advantage of this decomposition is that each term is
individually gauge invariant. So it has been used in theoretical studies of the nucleon spin in recent
years. However we had pointed out from the very beginning that, excepting the quark spin term, the
individual term does not satisfy the $SU(2)$ AM algebra. In addition, the term for the
gluon total AM has not been decomposed further into spin and orbital parts~\cite{ref10}:
The Poynting vector $\bf{E}\times\bf{B}$ is usually called momentum density of the em field, but the
$\bf{x}\times(\bf{E}\times\bf{B})$ term in Eq. (\ref{gid}) includes both OAM and
spin densities. When interactions are present, the resulting ``total AM'' does not
satisfy the $SU(2)$ AM algebra. So the third term in this gauge invariant decomposition
is strictly speaking not a proper AM operator.

In order to obtain a decomposition in which the individual term is not only gauge invariant but also
satisfies the Poincar$\acute{e}$ algebra and keeps the standard physical meaning as much as possible,
we proposed a new gauge invariant canonical decomposition in 2008~\cite{ref11},
\begin{eqnarray}
\bf{J}&=&\int \,d^3 x \psi ^\dagger \frac 12 \bf{\Sigma} \psi
+ \int d^3x \,\psi ^\dagger \bf{x} \times\frac 1i \bf{D}_{{\rm pure}} \psi \nonumber \\
&+& \int d^3x \,\bf {E} \times \bf {A}_{{\rm phys}}+ \int d^3x \,E^i\bf{x}\times \bf{\nabla} A_{{\rm phys}}^i.
\label{gidJ}
\end{eqnarray}
Here $\bf{D}_{{\rm pure}} = \bf{\nabla} - ie\bf{A}_{{\rm pure}}.$
So $-i\bf{D}_{{\rm pure}}$ is the gauge invariant version of the canonical momentum that reduces
to the standard canonical momentum when $\bf{A}_{{\rm pure}} = 0$. The three components
of $-i\bf{D}_{{\rm pure}}$ commute with each other, the same as the canonical momentum.
The second term satisfies the standard OAM algebra. The commutators between this new momentum and
OAM is the same as those of canonical ones in the Poincar$\acute{e}$ algebra.
Due to these properties they are the gauge invariant canonical momentum and OAM.
We also obtain a corresponding decomposition of the total momentum,
\begin{equation}
\bf{P} = \int d^3x \,\psi^{\dag}\frac{\bf{D}_{{\rm pure}}}{i}\psi
+ \int d^3x \,E^i\bf{\nabla}A^i_{{\rm phys}}.
\label{gidP}
\end{equation}
The momenta and the OAMs in Eqs. (\ref{gidJ}) and (\ref{gidP}) keep the standard relation 
$\bm{\ell}=\bf{x}\times\bf{p}$ in the integrand. In contrast, the gauge invariant
decomposition (\ref{gid}) gives instead the density $\bf{x}\times\bf{p} = \bf{j} = \bm{\ell}+\bf{s}$,
where $\bf{p} = \bf{E}\times\bf{B}$ is the Poynting vector. It is not hard to check that the
individual terms in the last two decompositions, Eqs. (\ref{gidJ}, \ref{gidP}), are all gauge invariant
and satisfy the canonical momentum and AM quantization rule. Complications might arise if one uses the
helicity representation for massless bosons, here photons or gluons, due to the transversality of
$\bf{A}_{{\rm phys}}$~\cite{ref12,ref13}.

The decomposition (\ref{gidJ}) reduces to the JM decomposition (\ref{JM}) in the Coulomb gauge. So the JM
operators are perfectly acceptable when used in the Coulomb gauge. Likewise, the usual gauge-dependent
canonical operators for electron momentum and OAM when used in the Coulomb gauge give
the same results as gauge invariant operators. The use of Coulomb gauge explains why the many results
calculated with ``gauge dependent'' canonical momentum and OAM are consistent with
measurements. If one goes beyond the Coulomb gauge, however, one can get wrong results~\cite{ref14}.

\section{Discussion}
\label{sec:4}

Ji tried to use his gauge invariant mechanical or kinematic momentum to explain why the gauge dependent
canonical momentum should be used in quantum mechanics to compare theory with experiment~\cite{ref15}.
He argued that the vector potential is $(\alpha)^2_{{\rm em}}$ suppressed when
compared to the canonical momentum, and so the latter is approximately gauge invariant. Gauge invariance
is an exact symmetry. There is only gauge invariance or non-invariance; approximate gauge invariance is
still gauge non-invariance. He also misinterpreted Feynman's idea about the quantum mechanical momentum.
Feynman wrote, {\it I'd like to make a brief digression to show you what this is all about - why there
must be something like Eq.(21.15) (${\bf p}\,\, momentum = m{\bf v} + \,q{\bf A}$) in the quantum
mechanics}~\cite{ref8}. Here Feynman clearly asserted that only the {\bf p}-momentum (canonical) and not
the kinematical m{\bf v}-momentum should be quantized in quantum mechanics. Because canonical quantization
quantizes the canonical momentum and coordinate, the three components of the mechanical (kinematical) momentum
do not commute; they cannot be diagonalized simultaneously. They then do not make up a complete commuting
set to give a momentum representation that is completely equivalent to the coordinate representation based
on a position operator whose components commute.

Up to now nobody has solved the eigenvalue equation for the kinematic momentum operator. Even if it would
be solved, the eigenfunctions will not be the plane waves now used in quantum physics. The resulting
x$\leftrightarrow$p representation transformations will not be the usual Fourier transformations that
connect a flat position space without interactions to a flat momentum space without interaction. With
the kinematic momentum operator for different physical systems containing different vector gauge
potentials, one will not have a universal momentum representation for all physical systems. It would
be very confusing indeed if the very meaning of physical momentum changes as soon as the interaction
changes. The quark ``OAM'' introduced by Ji in the gauge invariant decomposition (\ref{gid})
has similar problems.

There is still no consensus about the decomposition of the total momentum and AM of gauge systems.
One critique to the decompositions (\ref{gidJ}) and (\ref{gidP}) is that it uses non-local operators
$A^{\mu}_{{\rm phys}}$ and $A^{\mu}_{{\rm pure}}$. We believe that such non-local operators are
perfectly acceptable. First, the decomposition in the Coulomb gauge gives standard results. Second, the
renowned Aharonov-Bohm (A-B) effect is a non-local effect that is perfectly described by the decomposition of
the gauge potential into $A^{\mu}_{{\rm phys}}$ and $A^{\mu}_{{\rm pure}}$~\cite{ref16}. Finally, non-local
operators are already used in nucleon structure studies. For example, parton distributions all come from
non-local operators.

A second critique of the decompositions (\ref{gidJ}) and (\ref{gidP}) is that they are not Lorentz
covariant. This criticism is due to a misunderstanding of Lorentz transformations. The physical 4-coordinate
and 4-momentum can be measured in different Lorentz frames between which they are known to transform with
the well-known homogeneous Lorentz transformation law. The 4-vector potentials $A^{\mu}$ is not measurable
because they are not uniquely defined due to the gauge degree of freedom. Gauge invariance requires
that any gauge fixing must be Lorentz frame independent, otherwise the two fundamental principles, gauge
invariance and Lorentz covariance, will interfere with each other. Such a Lorentz frame independence can
be realized by including the gauge degree of freedom in the Lorentz transformation: In general, the
4-vector potential transforms with the non-homogeneous Lorentz transformation law,
\begin{equation}
A'_{\mu}(x') = \Lambda_{\mu}^{\nu}A_{\nu}(x) + \partial'_{\mu}\omega(x').
\end{equation}
Even the Lorentz gauge fixing, which is usually assumed to be transformed with homogeneous Lorentz
transformation law, can actually contain the non-homogeneous term $\partial'_{\mu}\omega(x')$ due to the
residual gauge degree of freedom. By means of this gauge degree of freedom, one can retain the Lorentz
covariance of any gauge fixing in every Lorentz frame.

A typical example is the Coulomb gauge fixing $\nabla\cdot\bf{A(x)}=0$. In this gauge, the usual homogeneous
Lorentz transformation will add an unphysical pure gauge part to the vector potential in the new Lorentz
frame. The non-homogeneous term $\partial'_{\mu}\omega(x')$ gives a second contribution that exactly cancels
the unphysical pure gauge part to enforce the transversality condition $\nabla'\cdot\bf{A'(x')}=0$ in the new
Lorentz frame. Such a non-homogeneous Lorentz transformation had been used in well known text books for quite
some time~\cite{ref17,ref18}.

A third critique is that the decompositions (\ref{gidJ}) and (\ref{gidP}) are not unique; infinitely
many other gauge invariant decompositions can be constructed by using the so-called gauge invariant
extension. Gauge invariance is a necessary condition for an operator to be measurable, but insufficient 
to fix the decomposition. One has to add a physical condition to fix the decomposition. We proposed to do 
this with the transverse wave condition $\nabla\cdot\bf{A_{{\rm phy}}}=0$. This photon condition and 
$F^{\mu,\nu}_{{\rm pure}}=0$ guarantee that ${\bf A_{{\rm phy}}}$ of Eq.(\ref{Feq}) has a unique 
solution. The photon condition is just the well known Helmholtz theorem.
Because an em wave is transverse, its physical vector potential has only two dynamical degrees of
freedom that reside in the 2D transverse space. Coulomb gauge fixing thus includes only
the physical degrees of freedom; there is no unphysical state in the Hilbert space.
The optics community confirmed these results both theoretically and experimentally~\cite{ref12,ref13}. 
Our treatment is also consistent with Dirac's description of gauge invariance in QED~\cite{ref5}.

The extension of the transverse wave condition to non-Abelian cases needs further study. We only proved
that the perturbative solution of this separation is unique. For strong gauge transformations involving
different winding number vacua, there might be additional complications, such as the Gribov
ambiguity~\cite{ref19}. In other physical condition choices,
the pure gauge part usually does not separate out
completely from the physical part. For example, the light-cone gauge still contains a residual gauge
degree of freedom. See the recent comprehensive review \cite{ref17} for more information.

Perhaps the best argument against Ji's idea that nonlocal operators are not measurable is provided by the
photon spin. Because photons travel with light speed, the measurable photon spin operator is the
helicity $\bf{S}\bm{\cdot\kappa}, \; \bm{\kappa} = {\bf p}/|p|$, where the projection into the momentum
direction involves a nonlocal operation in ordinary space.
The complete spin operator ${\bf S}' = \bm{\kappa}({\bf S}\cdot\bm{\kappa})$ is Abelian:
$\bf{S}' \times \bf{S'} = 0$, because projected values along the single momentum direction commute.
Our spin operator is based on the density $\bf{E_{\perp}}\times\bf{A_{\perp}}$~\cite{ref11,ref20} that
points in the same momentum direction and is also nonlocal. The nonlocality of  ${\bf S}'$ has not
prevented the helicity from being measured in atomic, nuclear, particle and optics physics.

There are still some issues concerning the photon OAM operator: The operator $\bf{L}'' = \bf{J - S'}$
proposed in optics~\cite{ref12} does not satisfy the usual $SO(3)$ AM algebra. Yet its expectation
value in light wave has been measured~\cite{ref12}. Our OAM is based instead on the OAM density
$-\bf{E^j_{\perp}}(i{\bf L})\bf{A_{\perp\,j}}$, where ${\bf L} = {\bf x} \times {\bf p}$ is the usual
SO(3) OAM operator and ${\bf p} = -i\bm{\nabla}$~\cite{ref20}. Two other terms in our
decomposition, one each from the spin and OAM densities, cancel out. This is why our operator
${\bf S}'$ does not contain any transverse part and is Abelian. There is a need to understand the
difference between these two approaches.

Some final words on the momentum operator: Ji\cite{ref15} and Wakamatsu~\cite{ref21} have insisted that in 
deep inelastic scattering, the measured quark momentum is the mechanical or kinematical momentum because 
of its appearance in the classical Lorentz equation. We agree that this classical mechanical (Feynman's 
m$\bf{v}$) momentum is indeed measurable in classical physics. The situation in quantum mechanics is 
different, however: One can only measure the quantum canonical momentum because only this operator has 
components that commute, and are therefore simultaneously measurable. The same conclusion also follows 
from studies of the second moment of the quark parton distribution that is related to the matrix element
$\langle PS|p^+ - gA^+|PS\rangle$ on the light cone. This form was initially misinterpreted as the ME
of the light-cone mechanical momentum ${\bf p} - g{\bf A}$. The misunderstanding was finally
corrected by Ji {\it et al.} who showed that in the infinite momentum frame, the vector potential $\bf{A}$
does not include the physical transverse part~\cite{ref22}. A study of the gauge link in the
collinear approximation also shows that only the longitudinal component of the gauge potential
(which is a nonphysical pure gauge contribution) is included in the gauge link.  Both results confirm
that the measured quark momentum distribution is the ME of $p^+ - gA^+_{{\rm pure}}$, the gauge invariant
{\it canonical} momentum ${\bf p} - g{\bf A}_{{\rm pure}}$ in the infinite momentum frame, in exact
agreement with our decomposition (\ref{gidP}). Our proposed parton distributions~\cite{ref11} thus
express better the physics of quark momentum and gluon helicity parton distribution.

This work is supported by the NSFC grant 1175215, 11175088, 11035006.

%\subsection{Subsection title}
%\label{sec:3}
%as required. Don't forget to give each section
%and subsection a unique label (see Sect.~\ref{sec:1}).
%\paragraph{Paragraph headings} Use paragraph headings as needed.
%\begin{equation}
%\end{equation}


\begin{thebibliography}{99}


%\bibitem[Author I(1999)]{Ref1}
%Author I (year) Article title. Journal Title-Abbreviated Vol: pp--pp
%\bibitem[Author and Smith(2001)]{Ref2}
%Author I, Smith J (year) Book title. Publisher, Place, pp numbers
% Format for proceedings
%\bibitem[Author and Smith(2003)]{Ref3}
%Author I, Smith J (year) Paper title. In: Editor, A. (ed.) Proceedings
%Title, Location, Date, pages. Publisher, Place
% etc


%\bibitem[Manohar A (1990)]{ref2}
%Manohar A (1990) Parton distributions from an operator viewpoint. Phys. Rev. Lett. 65:2511-2514
\bibitem[Zeng J Y(2000)]{ref1}
Zeng J Y(2000) Quantum mechanics. Academy, Beijing
\bibitem[Gottfried K (1966)]{ref2}
Gottfried K (1966) Quantum mechanics. Benjamin, New York pp 103--16
\bibitem[Berestetskii V B(1982)]{ref3}
Berestetskii V B, Lifshitz E M, Pitaevskii L P(1982) Quantum electrodynamics. Pergamon, Oxford;
Cohen-Tannoudji C, Dupont-Roc J Grynberg G(1989) Photons and atoms. Wiley, New York
\bibitem[Ji X(2010)]{ref4}
Ji X(2010) Comment on Spin and OAM in gauge theories. Phys. Rev. Lett. 104:039101;
106:259101

\bibitem[Dirac P A M(1955)]{ref5}
Dirac P A M(1955) Gauge-invariant formulation of quantum electrodynamics. Can. J. Phys. 33:650-660

%\bibitem[Bashinsky S V, Jaffe R L(1999)]{ref6}
%Bashinsky S V, Jaffe R L(1999) Quark and gluon OAM and spin in hard process.
%Nucl. Phys. B536:303-317

\bibitem[Jaffe R L, Manohar A(1990)]{ref7}
Jaffe R L, Manohar A(1990) The $g_1$ problem:Deep inelastic electron scattering and the spin of proton.
Nucl. Phys. B337:509-546

\bibitem[Feynman R P(1963)]{ref8}
Feynman R P, Leighton R B, Sands M(1963) The Feynman Lectures on physics, Vol.III. Addison-Wesley, Reading,
Ma, Menlo Park, California.

\bibitem[Chen X S, Wang F(1997)]{ref9}
Chen X S, Wang F(1997) The Problem of gauge invariance in the current study of nucleon spin.
Commun. Theor. Phys. 27:121-124;
Ji X(1997) Gauge-invariant decomposition of nucleon spin. Phys. Rev. Lett. 78:610-613

\bibitem[Chen X S, Wang F(1998)]{ref10}
Chen X S, Wang F(1998) Gauge invaiance and hadron structure. arXiv:hep-ph/9802346[hep-ph]

\bibitem[Chen X S(2008)]{ref11}
Chen X S, L\"u X F, Sun W M, Wang F, Goldman T(2008) Spin and OAM in gauge theories:
Nucleon spin structure and multipole radiation revisited. Phys. Rev. Lett., 100:232002-1-4; Do gluons carry half of the nucleon momentum? 103:062001-1-4

\bibitem[Bliokh K Y(2014)]{ref12}
Bliokh K Y, Dressel J, Nori F(2014) Conservation of the spin and OAMs in electromagneytism.
New J. Phys.16:093037;arXiv:1404.5486[physics.optics]

\bibitem[Beth R A (1936)]{ref13}
Beth R A (1936) Mechanical detection and measurement of the AM of light. Phys. Rev. 50:115-125;
Allen L et al (1992) OAM of light and the transformation of Laguerre-Gaussian laser modes.
Phys. Rev. A45: 8185-8189;
Van Enk S J, Nienhuis G(1994)  Commutation rules and eigenvalues of spin and OAM of radiation fields.
J. Modern Optics. 41:963-977

\bibitem[Sun W M(2010)]{ref14}
Sun W M, Chen X S, L\"u X F, Wang F(2010) Gauge invariant hydrongen-atom Hamiltonian.
Phys. Rev. A82:012107-1-5; Goldman T(1977) Gauge invariance time-dependent Foldy-Wouthuysen transformation, and the
Pauli Hamiltonian. Phys. Rev. D15:1063-1067; Kobe D H, Yang K H(1980) Gauge-invariant non-relativistic limit of an electron in a time-dependent electromagnetic field. J Phy. A13:3171-3185

\bibitem[Ji X(2012)]{ref15}
Ji X D, Xu Y, Zhao Y(2012) Gluon spin, canonical momentum, and gauge symmetry. JHEP 1208:082

\bibitem[Li J F(2012)]{ref16}
Li J F, Jiang Y, Sun W M, Zong H S, Wang F(2012) New application of decomposition of U(1) gauge potential:Aharonov-Bohm effect and Andereson-Higgs mechanism. Mod. Phys. Lett. B26:1250124-1-9

\bibitem[Leader E, Lorce C(2013)]{ref17}
Leader E, Lorce C(2013) The AM contraversy: What's it all about and does it matter?
arrXiv:1309.4235[hep-ph]

\bibitem[Bjorken J D(1965)]{ref18}
Bjorken J D, Drell S D(1965) Relativistic quantum fields. McGraw-Hill, New York;
Weinberg S(1996) The Quantum Theory of Fields, Vol I. Cambridge, U.K.

\bibitem[Zhang P M(2012)]{ref19}
Zhang P M, Pak D G(2012) On gauge invariant nucleon spin decomposition. Eur. Phys. J. A48:91-95;
Chen X S, Sun W M, Wang F, Goldman T(2011) Proper identification of the gluon spin. Phys. Lett. B700:21-24

\bibitem[Wong C W (2010)]{ref20}
Wong C W, Wang F, Sun W M, L\"u X F (2010) Gauge-invariant momentum and angular momentum operators in quantum electrodynamics and chromodynamics. arXiv:1010.4336[hep-ph]

\bibitem[Wakamatsu M(2011)]{ref21}
Wakamatsu M Gauge- and frame-independent decomposition of nucleon spin, Phys. Rev. D83:014012

\bibitem[Ji X (2013)]{ref22}
Ji X, Zhang J H, Zhao Y(2013) Physics of Gluon helicity contribution to proton spin. Phys. Rev. Lett. 111:112002



\end{thebibliography}
\end{document}